\documentstyle[12pt,epsfig]{article}

\newlength{\dinwidth}
\newlength{\dinmargin}
\def\lapproxeq{\lower .7ex\hbox{$\;\stackrel{\textstyle
<}{\sim}\;$}}
\def\gapproxeq{\lower .7ex\hbox{$\;\stackrel{\textstyle
>}{\sim}\;$}}

\def\be{\begin{equation}}
\def\ee{\end{equation}}
\def\bea{\begin{eqnarray}}
\def\eea{\end{eqnarray}}

\def\lapproxeq{\lower .7ex\hbox{$\;\stackrel{\textstyle
<}{\sim}\;$}}
\def\gapproxeq{\lower .7ex\hbox{$\;\stackrel{\textstyle
>}{\sim}\;$}}
\def\be{\begin{equation}}
\def\ee{\end{equation}}
\def\bea{\begin{eqnarray}}
\def\eea{\end{eqnarray}}

\def\ra{\longrightarrow}

\def\bb{b\bar{b}}

\def\ra{ \rightarrow }

\begin{document}
\titlepage
\begin{flushright}
IPPP/02/44 \\
DCPT/02/88 \\
24 September 2002 \\
\end{flushright}

\vspace*{2cm}

\begin{center}
{\Large \bf Diffractive Higgs Production: Myths and Reality}

\vspace*{1cm}
\textsc{V.A. Khoze$^{a,b}$, A.D. Martin$^a$ and M.G. Ryskin$^{a,b}$} \\

\vspace*{0.5cm} $^a$ Institute for Particle Physics Phenomenology,
University of Durham, DH1 3LE, UK \\
$^b$ Petersburg Nuclear Physics Institute, Gatchina,
St.~Petersburg, 188300, Russia \\
\end{center}

\vspace*{1cm}

\begin{abstract}
A critical comparison is made between recent predictions of the
cross sections for diffractive Higgs production at the Tevatron
and the LHC. We show that the huge spread of the predictions
arises either because different diffractive processes are studied
or because important effects are overlooked. Exclusive production
offers a reliable, viable Higgs signal at the LHC provided that
proton taggers are installed.
\end{abstract}

\section{Introduction}

Diffractive Higgs production may play an important role in
identifying and studying a $C$-~and $P$-even, light Higgs boson at
the LHC, see, for example, Ref.~\cite{DKMOR}. There exist a wide
range of predictions from a variety of models for the cross
section for diffractive Higgs production, which have yielded
answers ranging over many orders of magnitude. One unfortunate
consequence is to discredit diffractive Higgs production as a
possible way to identify a Higgs boson. Here we emphasize that the
huge spread of predictions is {\em either} because different
diffractive processes have been considered {\em or} because
important effects have been neglected. One of the aims of this
note is to guide the reader through the plethora of predictions,
making critical comparisons between the different approaches
wherever possible.

Let us consider a light Higgs boson (with mass less than 130~GeV)
with the dominant $H\ra\bb$ decay. From an observational point of
view, it is convenient to discuss three different diffractive
production mechanisms, where we will use a +~sign to indicate the
presence of a rapidity gap.

\begin{itemize}
\item[(a)] {\bf Exclusive production}:\ \ $pp\ra p + H + p$\\
If the outgoing protons are tagged, this process has the advantage
that the Higgs mass may be measured in two independent ways;
first, by the missing mass to the outgoing protons and, second, by
the $H\ra\bb$ decay. So the signal must satisfy $M_{\rm miss} =
M_{\bb}$, with allowance for experimental resolution\footnote{This
way to identify a light Higgs boson in Run~II of the Tevatron was
proposed in Ref.~\cite{AR}. The experimental issues concerning the
LHC measurements are covered in \cite{DKMOR}.}. Moreover, the
$\bb$ background is suppressed by a spin ($J_z=0$) selection rule,
which leads to a favourable signal-to-background ratio.

\item[(b)] {\bf Inclusive production}:\ \ $pp\ra X + H + Y$\\
The advantage is a much larger cross section. However, there is no
spin selection rule to suppress the $\bb$ background, and the
signal-to-background ratio is unfavourable. Moreover, the accuracy
of the Higgs mass determination is worse, as $M_{\rm miss}$ is not
applicable.

\item[(c)] {\bf Central inelastic production}:\ \ $pp\ra p + (HX) + p$\\
There is additional radiation accompanying the Higgs in the
central region, which is separated from the outgoing protons by
rapidity gaps. Although this mechanism is often used for
predictions, it has, in our view, no special advantages for Higgs
detection.
\end{itemize}

We may regard each large rapidity gap as being generated by an
effective Pomeron exchange. It may be {\em either} a QCD Pomeron,
which at lowest order is a gluon--gluon state, {\em or} a
phenomenological Pomeron with parameters fixed by data. The above
information is summarised in Fig.~1, together with a leading order
QCD diagram of each process.

\vspace{0ex}
\begin{figure}[h]
\begin{center}
\epsfig{figure=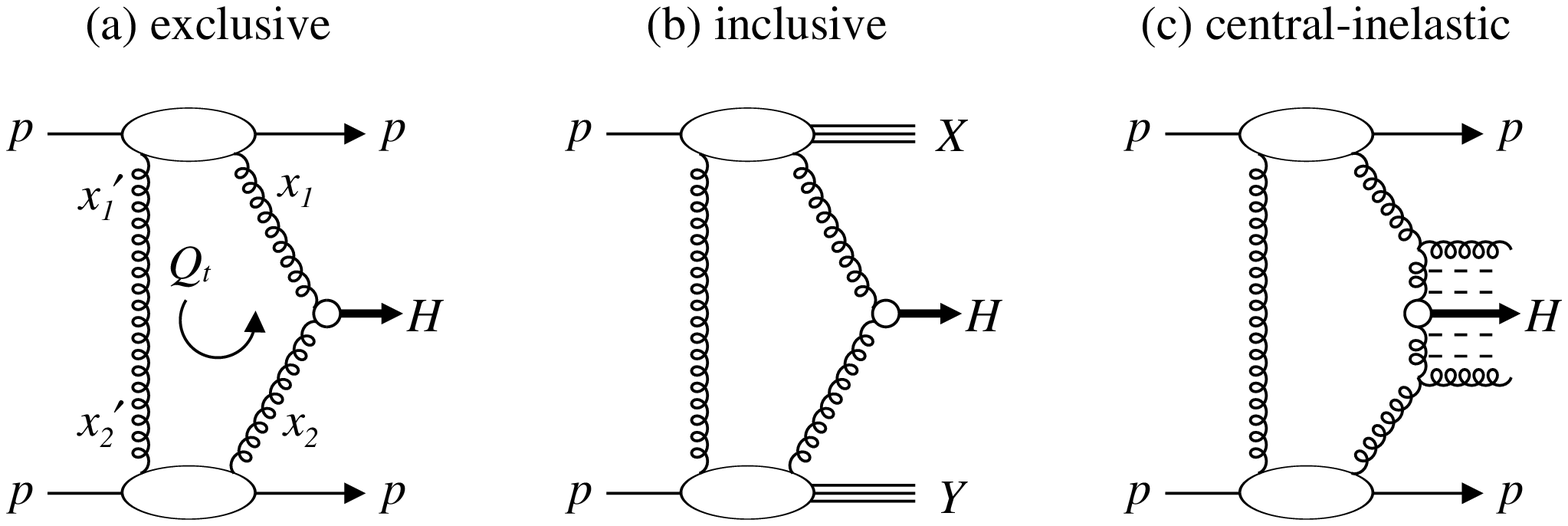,height=4.5cm,width=15.2cm}
\end{center}\end{figure}
\vspace{-1.35cm}
\begin{table}[h]
\begin{center}
\begin{tabular}{@{}  c  @{\hspace{1.5cm}}  c  @{\hspace{1.6cm}}  c  @{\hspace{2mm}}}
$M_{\rm miss}=M_H$  &  no $M_{\rm miss}$  &  $M_{\rm miss}>M_H$ \\
$J_z=0$ rule for background  &  no rule  &  no rule \\
$S/B\sim3$  &  $S/B\sim0.01$  &  $S/B\sim0.001$ \\
pile-up may be overcome  & pile-up problems\ \ & pile-up may be
overcome
\end{tabular}
\end{center}
\end{table}\vspace*{-7ex}
\begin{figure}[!h]
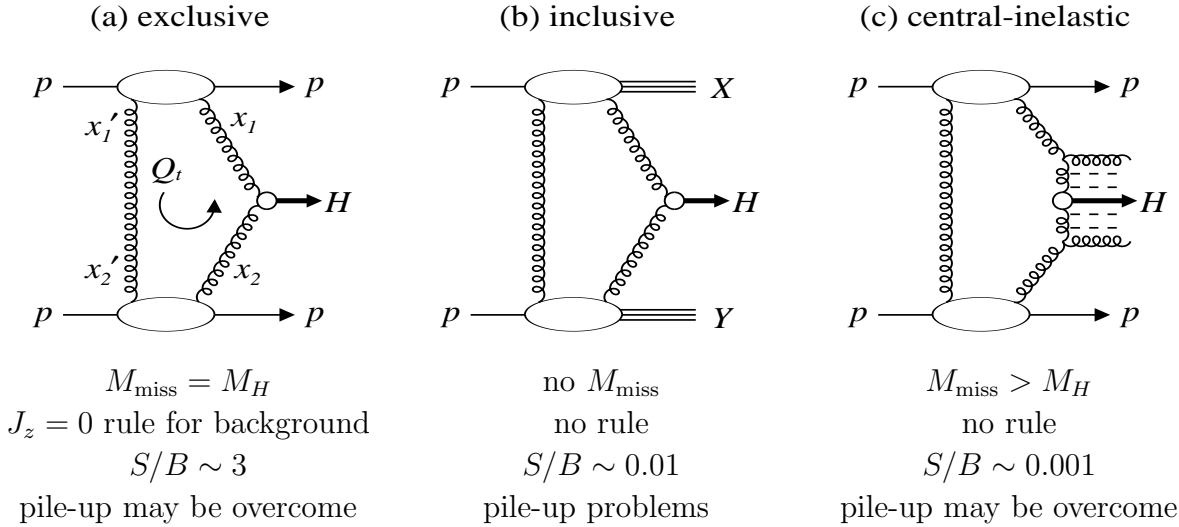

\caption{\small  Different processes for double-diffractive Higgs
production in $pp$ collisions in terms of perturbative QCD. The
signal-to-background ratios, $S/B$, are obtained using the mass
resolutions, $\Delta M_{\rm miss} = 1$~GeV and $\Delta M_{\bb} =
10$~GeV expected for experiments at the LHC~\cite{DKMOR}. Pile-up
refers to the multiple interactions per bunch crossing at the
LHC.}
\end{figure}

\newpage

Recall that, at medium and high luminosity at the LHC, the
recorded events will be plagued by overlap interactions in the
same bunch crossing. For example, at the medium luminosity of
$10^{33}$~cm$^{-2}$s$^{-1}$, an average of 2.3 inelastic events
are expected for each bunch crossing. Hence the rapidity gaps
occurring in one interaction may be populated by particles created
in an accompanying interaction. It is, however, possible to use
detector information to locate the vertices of the individual
interactions and, in principle, to identify hard scattering events
with rapidity gaps. For the exclusive and central inelastic
processes of Figs.~1(a) and 1(c), the use of proton taggers makes
it much more reliable to select the rapidity gap
events\footnote{In addition to helping to select events with
rapidity gaps, it may be possible to use the proton taggers to
measure the approximate position of the vertex of the event,
although the accuracy is expected to be $\pm$3--5~cm at best.}.

We note that the significance of a signal is only increased if the
background-to-signal ratio is decreased by more than the signal.
This does not happen for either (b)~inclusive production or
(c)~central-inelastic production. Indeed, the corresponding
effective (Pomeron--Pomeron) luminosities are orders of magnitude
smaller than the gluon--gluon luminosity which governs the
conventional totally inelastic Higgs production signal, while the
background-to-signal ratio is decreased by at most a factor of two
due to the lower hadronic multiplicity.

In principle, factorization does not hold for the diffractive
processes; it can only occur by chance. It is spoilt by the
Pomeron and Reggeon cut contributions, and by QCD radiation.
Moreover, the comparison~\cite{CDFfac} of CDF dijet and HERA
diffractive data demonstrates a strong violation of
factorization~\cite{COMPAT}. Nevertheless, the existing approaches
may be classified as factorizable or non-factorizable. Some
authors \cite{PP1,PP2,BDPR,AF} use factorization \`a la
Ingelman-Schlein model~\cite{IS} and then introduce a normalising
factor to account for the non-factorizable nature of the process.

\section{Survival probability of the rapidity gaps}

The cross sections for processes with rapidity gaps are reduced by
the probabilities of the gaps not to be populated by, first, the
gluon radiation associated with a QCD Pomeron and/or the hard
$gg\ra H$ subprocess and, second, by secondaries produced in the
soft rescattering of the spectator partons. We denote these
survival probabilities by $T^2$ and $S^2$ respectively. The
probability amplitude $T$, not to radiate, can be calculated using
perturbative QCD. The expression for $T$ has the familiar
double-logarithmic Sudakov form. In fact it is possible to include
the next-to-leading (single log) corrections in $T$~\cite{KMR}.
Note that the amplitude $T$ plays a crucial role in providing the
infrared convergence of the loop integral over the $t$ channel
gluon transverse momentum $Q_t$. On the other hand the survival
factor, $S^2$, to soft rescattering cannot be calculated
perturbatively. The presence, and the value, of $S^2$ has been
checked experimentally by comparing the diffractive cross section
in deep inelastic reactions at HERA (where $S$ is close to 1, due
to the absence of soft spectators in the virtual photon) with the
cross section of diffractive dijet production at the Tevatron, for
which it turns out that $S^2 \sim 0.1$~\cite{CDFfac}. Theoretical
predictions of the survival factor, $S^2$, can be found in
Refs.~\cite{KMRsoft,KKMR,GLM}. Note that the factor $S^2$ is not a
universal number. It takes account of soft rescattering in both
the initial and final state interactions. Therefore the value of
$S^2$ depends on the initial energy and the particular final
state. Clearly, the presence of such a factor violates
factorization~\cite{KMRcalabria}.

It is informative to digress for a moment and to note that the
large rapidity gaps may also be produced by electroweak
interactions, as well as by QCD. Exclusive Higgs production can
proceed by $\gamma\gamma$ fusion\footnote{The known $\gamma\gamma$
fusion process provides a lower limit for exclusive Higgs
production; see, for example, \cite{INC} and references therein.},
while the inclusive reaction can proceed via weak boson fusion. As
the photon and the $W$ boson are point-like colour singlets, there
is no $T^2$ factor in these reactions. However, we still have to
allow for the survival factor, $S^2$, to soft rescattering. The
process $\gamma\gamma\ra H$ is dominated by photons with very
small transverse momenta, which corresponds to the interaction
occurring mainly at large impact parameters. There are two
consequences. First, the factor $S^2\simeq1$ (or, to be precise,
0.86 at the LHC, see~\cite{INC}) and, second, there is almost no
interference between the QCD and $\gamma$ exchange amplitudes. In
the case of weak boson fusion the interference is also suppressed,
but now due to the much larger transverse momentum (about $M_W/2$)
transferred to the Higgs boson. The survival factor $S^2$ is
comparable to that for the QCD-induced reaction; in fact a bit
larger, since the $W$~bosons are emitted from valence quarks which
are more concentrated in the component of the proton wave function
which has smaller absorption~\cite{KMR, KKMR, INC}.

\section{Exclusive diffractive Higgs production}

The first QCD-based calculation of double-diffractive Higgs
production was performed by Bialas and Landshoff~\cite{BL}. In
terms of QCD the Higgs boson is produced by gluon--gluon fusion.
The colour of these $t$~channel gluons is screened, at leading
order, by an accompanying $t$ channel gluon exchange between the
incoming protons, as shown in the diagrams in Fig.~1. The
screening is necessary as colour cannot be transferred across a
rapidity gap; colour flow would populate the gap with secondaries
via hadronization. Thus we need two-gluon colour-singlet exchange
across the gaps. Effectively in this approach, we may regard the
Higgs as being produced by the fusion of two
`Low--Nussinov'-like~\cite{LN} di-gluon Pomerons. A more precise
understanding of this statement will emerge as we discuss the
different production mechanisms below.

Bialas and Landshoff~\cite{BL} did not include in the calculation
the probabilities $T^2$ and $S^2$ that the rapidity gaps survive
QCD radiation and soft rescattering respectively, although they
left open the possibility that extra radiation may result in an
inclusive process. However, they did not quantify this effect. A
number of authors have adopted a similar approach to calculate the
cross section for Higgs production with large rapidity gaps. Some
of the most recent calculations are listed in Table~1.

\begin{table}[h]
\begin{center}
\begin{tabular}{@{} l @{\ }|@{\ } c @{\ }| c | c | c @{\ }| c | c | l @{}}
\raisebox{-1.5ex}[0ex][1ex]{Reference}  &
\raisebox{-1.5ex}[0ex][1ex]{Process}  &
\multicolumn{2}{c}{Survival factor} \vline&
\raisebox{-1.5ex}[0ex][1ex]{Norm.}  &  \multicolumn{2}{c}{$\sigma_{\rm Higgs}$~(fb)}  \vline&  \raisebox{-1.5ex}[0ex][1ex]{Notes} \\
\cline{3-4}\cline{6-7} &  &  $T^2$  &  $S^2$  &  &  Teva.  &  LHC  &  \\
\hline%ROW 1 (TITLES) ENDS HERE---------------------
Cudell,  & excl  &  no  &  no  & $\sigma_{\rm tot}$  &  30
&  300  &  Overshoots CDF dijets \\  Hernandez~\cite{CH}  &  incl  &  &  &  &  200  &  1200  &   by 1000. \\
\hline%ROW 2 ENDS HERE---------------------
\raisebox{-1.5ex}[0ex][1ex]{Levin~\cite{L}} & excl & yes & yes &
$\sigma_{\rm tot}$  &
20  &    & Overshoots CDF dijets \\  &  incl  &  No DL  &&&  70  &  \raisebox{1.5ex}[0ex][1ex]{--}  &  by 300.  \\
\hline%ROW 3 ENDS HERE---------------------
\raisebox{-1.5ex}[0ex][1ex]{Khoze, Martin,} &
excl &  & &  pdf  &  0.2  &  3  &  \raisebox{-1.5ex}[0ex][1ex]{Uses skewed gluons.} \\
\raisebox{-1.5ex}[0ex][1ex]{Ryskin~\cite{INC}}  & incl & yes & yes
& pdf  &  1  &  40  &  \raisebox{-1.5ex}[0ex][1ex]{CDF dijets OK.}
\\  &  C.inel &  & &  &  $\sim0.03$  &  50 &
\\ \hline%ROW 4 ENDS HERE---------------------
Cox, Forshaw,  &  \raisebox{-1.5ex}[0ex][1ex]{C.inel} &
\raisebox{-1.5ex}[0ex][1ex]{$T\simeq1$}  & \raisebox{-1.5ex}[0ex][1ex]{norm}  &  CDF  & \raisebox{-1.5ex}[0ex][1ex]{0.02} & \raisebox{-1.5ex}[0ex][1ex]{6} & No LO, only NLO, QCD \\
Heinemann~\cite{PP1}  & & & &  dijet  & & &  i.e., no Fig.2(a), only 2(c).  \\
\hline
%ROW 5 ENDS HERE---------------------
Boonekamp,&&&&&&&
\\  De Roeck, & \raisebox{-1.5ex}[0ex][1ex]{C.inel}  &  \raisebox{-1.5ex}[0ex][1ex]{$T\simeq1$}  &
\raisebox{-1.5ex}[0ex][1ex]{norm}  &  CDF &
\raisebox{-1.5ex}[0ex][1ex]{2.7} &
 \raisebox{-1.5ex}[0ex][1ex]{320}
& No LO, only NLO, QCD.
 \\ Peschanski,   & & & & dijet  &  & & Assume
$S^2_{\rm CDF} = S^2_{\rm LHC}\,$.  \\
Royon~\cite{BDPR}  &&&&&&&  \\ \hline%ROW 6 ENDS HERE---------------------
Enberg, &&&&&&&   \\
Ingelman, & incl & \raisebox{-1.5ex}[0ex][1ex]{yes}
&  \raisebox{-1.5ex}[0ex][1ex]{yes}  &  \raisebox{-1.5ex}[0ex][1ex]{$F_2^{\rm Diff.}$}  &  \raisebox{-1.5ex}[0ex][1ex]{$<0.01$}  &  \raisebox{-1.5ex}[0ex][1ex]{0.2}  &  \raisebox{-1.5ex}[0ex][1ex]{No coherence.}\\
Kissavos,  &  C.inel  &  &  &  &  & &
\\ Timneanu~\cite{SCIH} &&&&&&&
\end{tabular}
\end{center}
\caption{\small Recent QCD-based calculations of the cross
section, $\sigma_{\rm Higgs}$, for exclusive, inclusive and
Central inelastic double-diffractive production of a Higgs boson
of mass about 120~GeV, at Tevatron and LHC energies. The Norm.
column indicates the way in which the various predicted cross
sections are normalised. $T^2$ and $S^2$ are the survival
probabilities of the rapidity gaps to QCD radiation associated
with the hard $gg\ra H$ subprocess and to soft rescattering,
respectively; ``norm'' in the $S^2$ column means that $S^2$ is
simply determined by normalising to CDF dijet data~\cite{CDFjj}.
The cross sections for central inelastic production (C.inel)
correspond to integrating up to $M_{\rm miss}=0.1\sqrt{s}$, where
$\sqrt{s}$ is the collider energy. Note that in Ref.~\cite{INC}
the C.inel cross section is 0.2~fb at the Tevatron, but this
includes the exclusive contribution. The LHC entry for Cox
et~al.~\cite{PP1} is obtained using $S^2=0.02$.}
\end{table}

We start the discussion of the results shown in Table~1 with the
calculation of the exclusive double-diffractive cross section by
Levin~\cite{L}. He assumed the survival probability to soft
rescattering $S^2=0.1$. To account for QCD radiation he multiplied
the final result by an effective $T^2$ factor which was estimated
phenomenologically assuming a Poisson probability
$\exp(-\bar{n})$, where $\bar{n}$ is the mean multiplicity of
mini-jets produced in hadron interactions with energy
$\sqrt{s}\simeq M_H$. This assumption overestimates the survival
factor $T^2$ in comparison with the perturbative QCD calculation,
since instead of getting the double-logarithmic Sudakov-like
suppression, his probability $\exp(-\bar{n})$ corresponds to a
single logarithm.

In the calculation by Cudell and Hernandez~\cite{CH}, both the
soft $S^2$ and hard $T^2$ survival factors were neglected. In
addition to the pure exclusive process, inclusive events where an
incoming proton dissociates into $N^*$ resonances were allowed, so
the predicted cross section becomes larger. A crucial point, both
in this calculation and in that of Levin, is the normalization of
the two-gluon exchange amplitude. Without the double-logarithmic
$T$~factor inside the loop integration over the gluon transverse
momentum $Q_t$, the integral is infrared divergent. To obtain a
finite result the authors have to choose an infrared cut-off or to
introduce a finite mass for the gluon. The value of the cut-off,
or mass, is tuned to reproduce the total $pp$ cross section,
$\sigma_{\rm tot}$, in terms of the Low--Nussinov two-gluon
Pomeron exchange. It has been noted~\cite{BER,KMRdijet} that the
use of such a prescription further overestimates the Higgs
production cross section. Indeed, in terms of $Q_t$ factorization,
the Higgs production forward amplitude is of the form
\begin{equation}
{\cal M}_{\rm Higgs} = A\pi^4\int\frac{dQ_t^2}{Q_t^4}\,
f_g(x_1,x_1^\prime,Q_t^2)\,f_g(x_2,x_2^\prime,Q_t^2)
\label{eq:M_higgs}
\end{equation}
where the factor $A$ represents the $gg\ra H$ vertex and
$f_g(x,x^\prime,Q_t^2)$ is the unintegrated skewed gluon density.
The unintegrated gluon density embodies the $T$ factor
\cite{KIMBER,KMR} which accounts for the fact that the gluon which
participates in the hard $gg\ra H$ subprocess remains untouched in
the evolution from $Q_t$ up to the hard scale, $\sim M_H/2$; this
hard scale is an implicit variable in the $f_g$
in~(\ref{eq:M_higgs}). Similarly, via the optical theorem, we may
express the total cross section in terms of two-gluon exchange
\begin{equation}
\sigma_{\rm tot} =
\frac{\pi^3}{2}\int\frac{dQ_t^2}{Q_t^4}\,f_g(x,x,Q_t^2)\,f_g(x,x,Q_t^2)
\label{eq:sigma_tot}
\end{equation}
where $x\lapproxeq2Q_t/\sqrt{s}$, as follows from the internal
kinematics of the process, and where the implicit scale in $f_g$
is now $\sim Q_t$. At first sight it appears that
(\ref{eq:sigma_tot}) will give a precise normalisation of the
Higgs cross section, via (\ref{eq:M_higgs}). However, in addition
to the different implicit scales, the typical values of $x$
sampled in (\ref{eq:sigma_tot}) are about two orders of magnitude
smaller than the values of $x_i\sim M_H/\sqrt{s}$ sampled in
(\ref{eq:M_higgs}). Since $f_g$ grows as $x$ decreases and since
$\sigma_{\rm Higgs}\propto|f_g|^4$, this normalisation
considerably overestimates the cross section for Higgs production.
Despite the fact that $S^2$ and $T^2$ factors were included in the
prediction of the cross section given in \cite{L}, the result is
close to that of \cite{CH}. One reason is that these small
survival factors are compensated by the use of a larger
value\footnote{It is argued in Ref.~\cite{L} that, instead of the
conventional $\alpha_S(M_H)$, a much larger QCD coupling (at low
scale, $\sim 1$~GeV) is to be taken. However, the high-order
evolution of the Higgs vertex confirms the former
choice~\cite{KMR}.} of $\alpha_S$ in the $gg\ra H$ vertex.

The reliability of the prediction of the diffractive production of
the Higgs boson can be checked experimentally by measuring the
much larger cross section for double-diffractive central
production of a pair of high $E_T$ jets~\cite{KMR,KMRdijet}. The
amplitude for this process has the same structure as
(\ref{eq:M_higgs}), with the $gg\ra H$ vertex replaced by the
matrix element of the $gg\ra gg$ subprocess. The original
calculation of dijet production was performed by Berera and
Collins~\cite{BC}. The result~\cite{BER}, with $S^2$ and $T^2$
factors neglected and normalised to $\sigma_{\rm tot}$, is about
5600~nb for CDF dijets at the Tevatron energy, in contrast with
the experimental upper limit of less than 3.7~nb at 95\%
confidence level~\cite{CDFjj}. The huge difference originates from
the product of three factors---the survival factors
$S^2\simeq0.05$--0.1 and $T^2\simeq0.1$--0.2 should be included in
the prediction, and the normalisation should be reduced by about a
factor of 10, since (\ref{eq:sigma_tot}) should be compared to
(\ref{eq:M_higgs}) at much lower $x$. Indeed Berera and
Collins~\cite{BC} had noted that the survival factors should be
computed before their leading order calculation is compared with
data. In fact, when account is taken of the survival factors, our
perturbative approach~\cite{KMR, KMRdijet} leads to the prediction
of about 1~nb \cite{Liverpool} for the exclusive production of
dijets corresponding to the kinematics\footnote{The accuracy of
the theoretical prediction for $E_T>7$~GeV jets at the Tevatron
energy is far from as good as the factor of 2 uncertainty claimed
for dijets of mass $M(jj)\sim M_H$ at the LHC. The contribution
from the low $Q_t$ domain is less under control for the CDF
kinematics of Ref.~\cite{CDFjj}.} of the CDF dijet
search~\cite{CDFjj}, which leads to a dijet bound of less than
3.7~nb.

Let us return to the discussion of the predictions listed in
Table~1. Since Cudell and Hernandez~\cite{CH} do not include the
$S^2$ and $T^2$ survival factors, and apply a $\sigma_{\rm tot}$
normalisation, we may expect that the dijet cross section would be
overestimated by a factor of about 1000. Levin~\cite{L} includes
estimates of the $S^2$ and $T^2$ factors and, following his
prescription, we would expect a dijet cross section of about
1000~nb, which is still much larger than the experimental limit.
There is no simple way of using these dijet overshoot factors to
correct the predictions for Higgs production given in
Refs.~\cite{L,CH}. We cannot simply scale down the predictions by
dividing by the overshoot factors. The correction factor has,
first, an energy dependence arising from the effective gluon
density normalised to $\sigma_{\rm tot}$ and, second, due to the
energy dependence of the soft survival factor $S^2$. Moreover, the
QCD radiative effects described by the $T$ factor depend strongly
on the hard scale, and are quite different for dijet production,
with jets of $E_T\sim7$--10~GeV, and Higgs production with scale
$M_H/2\sim60$~GeV.

Instead of fixing the normalisation of the prediction for
exclusive Higgs production by using $\sigma_{\rm tot}$, a more
reliable method is to use the gluon density given by global parton
analyses and to include the Sudakov-like survival factor
$T=\exp\left(-{\cal S}(Q_t^2,M_H^2)\right)$ inside the loop
integral over $Q_t$ in (\ref{eq:M_higgs})~\cite{KMRH}. This factor
provides the infrared stability of the integral, while the known
gluon distribution fixes the normalisation. More recently, the
method has been further improved~\cite{KMR,INC}. First, the skewed
effect is included (using the prescription of
Refs.~\cite{MR01,SGMR}), that is the effect due to unequal
longitudinal momentum fractions carried by the left and right $t$
channel gluons in Fig.~1(a): explicitly, we have
$(x_i^\prime\simeq Q_t/\sqrt{s})\ll(x_i\simeq M_H/\sqrt{s})$.
Second, the NLO corrections to the $gg\ra H$ vertex, and the
next-to-leading correction to the double-logarithmic $T$ factor
(that is the single log term in $T$), are included\footnote{Note
that the gluon with $x^\prime\simeq0$ is almost `at rest' and
practically does not radiate. Thus, the QCD radiation is
associated with the hard $gg\ra H$ subprocess.}. This is the
method used for obtaining the numbers quoted for the third
entry~\cite{INC} in Table~1.

The most delicate point, in the prediction of the cross section
for diffractive Higgs production, is the calculation of the
probability, $S^2$, that the rapidity gaps survive the soft
rescattering. $S^2$ cannot be determined using perturbative QCD
and non-perturbative techniques have to be applied. To improve the
accuracy of the prediction of $S^2$, a detailed
analysis\footnote{The data were analysed in terms of a two-channel
eikonal model, which also incorporated high mass diffraction and
$\pi$-loop insertions in the Pomeron trajectory (to describe
better the periphery of the proton).} of all available soft high
energy $pp$ and $p\bar{p}$ data was performed~\cite{KMRsoft}.
Using the results of this analysis it is possible to compute the
soft survival factor $S^2$ for a complete range of diffractive
processes. The factors for Higgs production are given in
Refs.~\cite{KMR,KMRsoft,INC}. For exclusive Higgs production at
the LHC the soft survival factor $S^2$ is found to be 0.02. After
all the above effects are included, the uncertainty in the
prediction of the cross section, $\sigma(pp\ra p+H+p)\simeq 3$~fb,
is estimated to be about a factor of two~\cite{DKMOR}.

\section{Inclusive diffractive Higgs production}

If we allow the protons to dissociate, but still keep the rapidity
gaps on either side of the produced Higgs boson, then we enlarge
the cross section by a factor of 3--10, depending on the range of
masses allowed for the dissociation~\cite{BL,CH,KMR,L,INC}. In
addition to the larger available phase space for inclusive
kinematics, also the gap survival factor is larger; in fact using
the formalism of Ref.~\cite{KMRsoft} we find $S^2_{\rm
incl}\simeq0.1$ at the LHC, while for exclusive and central
inelastic production we have $S^2\simeq0.02$. However, we lose all
the advantages of exclusive double-diffractive Higgs production.
In particular, we lose the good missing mass resolution provided
by the proton tagger, the equality $M_{\rm miss} = M_{\bb}$ from
the $H\ra\bb$ decay, and the suppression of the $\bb$ QCD
background and of the pile-up events. We therefore do not discuss
this process any further here.

\section{Central inelastic Higgs production}

So far we have considered processes where there are no secondaries
accompanying Higgs production in the central rapidity region. By
`central inelastic Higgs production' we mean that secondaries are
allowed in some central rapidity interval. Two contributions to
the process are sketched in Fig.~2.
\begin{figure}[h]
\begin{center}
\epsfig{figure=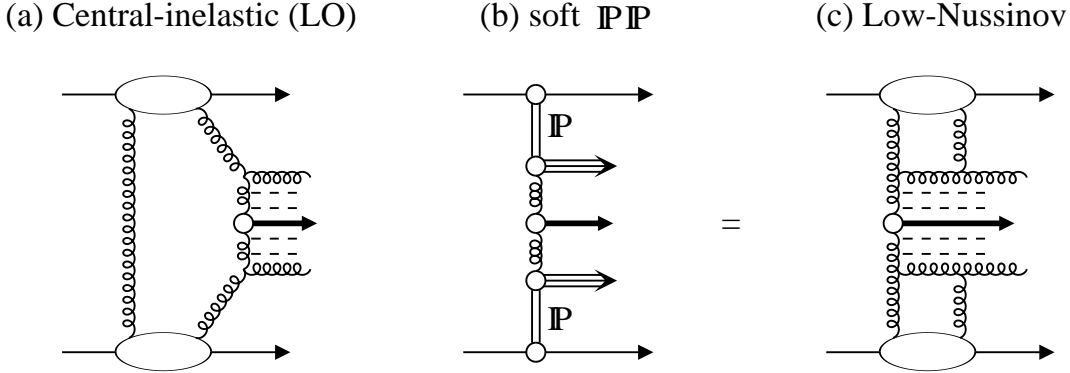,height=2in}
\end{center}
\vspace{-0.5cm}\caption{\small Central inelastic
double-diffractive production, in which the Higgs boson (shown by
the {\em bold} central arrow) is accompanied by gluon emission in
the central rapidity region. Diagram~(b) shows the contribution
from Pomeron--Pomeron inelastic collisions in which the Pomeron
remnants populate the central region. Diagram~(c) shows the
Pomeron--Pomeron production process from a QCD viewpoint, in which
each Pomeron is represented by two-gluon exchange. Diagrams~2(a)
and 2(b,c), respectively, may be regarded as lower and higher
order $\alpha_S$ contributions to central inelastic production.}
\end{figure} As we shall see in a moment, we may call diagrams~2(a)
and 2(b) lower and higher order $\alpha_S$ contributions
respectively. In fact, much attention is paid in the literature to
Higgs production in Pomeron--Pomeron inelastic collisions,
Fig.~2(b), which, in our notation, corresponds to the higher order
$\alpha_S$ contribution to central inelastic production. So we
discuss this first. The cross section for Higgs production by
Pomeron--Pomeron collisions is larger than for exclusive
diffractive production, but still much smaller than that for the
normal inclusive production, $pp\ra HX$. The expected
signal-to-background ratio is practically the same as for normal
inclusive production but at a lower energy, corresponding to the
Pomeron--Pomeron energy as measured by the missing mass method. We
have effectively degraded the LHC energy down to energies
comparable to the Tevatron! Of course, the luminosity of the LHC
is larger than that of the Tevatron. However, the effective
Pomeron--Pomeron luminosity contains its own small factors. The
only advantage\footnote{Also, $b$-tagging may be easier due to the
lower mean multiplicity of soft secondaries, see also~\cite{PP2}.}
of Higgs production by Pomeron--Pomeron inelastic collisions, in
comparison to normal inelastic production, at the LHC, is the
possibility to use proton taggers to avoid pile-up problems
(associated with multiple interactions in each bunch crossing).

Usually the cross section for central inelastic production is
estimated using the factorization hypothesis, \`a la
Ingleman--Schlein~\cite{IS}. To account for the probability $S^2$
that the rapidity gaps survive the soft rescattering (which
violates factorization), the predictions are normalised to the
observed rate of central inelastic double-diffractive dijet
production at the Tevatron~\cite{CDFjj}. From a QCD viewpoint, the
soft Pomeron--Pomeron interaction, Fig.~2(b), should be regarded
as Fig.~2(c) where the soft Pomerons are replaced by
(Low--Nussinov) two-gluon exchange. We note that Fig.~2(c)
contains an extra factor of $\alpha_S$, as compared to Fig.~2(a).
Of course, this coupling occurs at low scale, but nevertheless we
should not be surprised when we find that the contribution of
Pomeron--Pomeron collisions, Fig.~2(c), is less than that of
central inelastic production, Fig.~2(a). For example, in
Ref.~\cite{PP1} the cross section corresponding to Fig.~2(b,c) was
calculated using the H1~parameterisation of the Pomeron flux and
structure function~\cite{H1}. In the absence of dijet data at the
LHC energy, the LHC prediction of \cite{PP1} is presented without
accounting for the $S^2$ factor. If we multiply this result by the
same $S^2=0.02$ as in Refs.~\cite{KMRsoft,INC} we obtain
$\sigma({\rm C.inel})\simeq6$~fb, which is an order of magnitude
smaller than 50~fb -- the cross section corresponding to
Fig.~2(a).

%Suppose, for the moment, that we neglect the soft survival factor,
%$S^2$, and we compute the cross section corresponding to Fig.~2(c)
%using the Donnachie--Landshoff parameterization~\cite{DL} to
%calculate the Pomeron flux, and take a gluon structure function of
%the Pomeron consistent with the H1 analysis of deep inelastic
%diffractive data~\cite{H1}. Then, indeed, we do obtain a result
%smaller than that corresponding to Fig.~2(a).

Central inelastic production, Fig.1(c), may be regarded as
higher-order QCD radiative corrections to exclusive Higgs central
production (Fig.~1(a)). Allowing QCD radiation, in a central
rapidity interval around the Higgs boson, increases the
probability of gap survival, but weakens the potential of the $T$
factor to provide infrared convergence of the loop integral over
$Q_t$. The cross section is increased, with the extra contribution
coming from the low $Q_t$ region. In Ref.~\cite{INC}, results were
focused on central inelastic production allowing radiation only in
a relatively small central rapidity interval, $\delta\eta$. In
this case, the residual $T$ factor is still able to ensure
infrared convergence of the loop integral. If proton taggers are
installed, then the mass of the central system (that is the Higgs
plus accompanying radiation) can be measured by the missing mass
method. For the large masses, up to $M_{\rm miss} = 1.4$~TeV that
were considered in Ref.~\cite{BDPR}, the $T$ factor approaches
unity and almost any QCD radiation is allowed. In these
circumstances there is no convergence of the $Q_t$ integral, and
the only possibility is to normalise the prediction to
$\sigma_{\rm tot}$, recall (\ref{eq:M_higgs})
and~(\ref{eq:sigma_tot}). The typical values of $x$ sampled in
(\ref{eq:M_higgs}) are $x\sim M_H/\sqrt{s}\sim 0.01$ at the LHC.
To evaluate (\ref{eq:sigma_tot}) in a comparable $x$~domain, we
use the value of $\sigma_{\rm tot}$ at a much lower (CERN--ISR)
energy. Based on this normalisation, and including a soft survival
factor $S^2=0.02$, we predict a central inelastic cross section of
50~fb, which is to be compared to the 320~fb predicted in
Ref.~\cite{BDPR}. Strictly speaking, the 320~fb in \cite{BDPR} was
calculated for Pomeron--Pomeron inelastic collisions, Fig.~2(b).
For comparison, if for this latter process we were to use the
Donnachie--Landshoff~\cite{DL} parameterization for the Pomeron
flux, the Pomeron structure function as measured by \cite{H1}, and
the known soft survival factor $S^2$, we would obtain 1.7~fb at
the LHC. However, in \cite{BDPR} the cross section was normalized
using the CDF dijet data~\cite{CDFjj} at the Tevatron, for which
the dominant contribution comes from central inelastic diagrams of
the type Fig.~2(a); so comparison with 50~fb is more relevant.
Note that the dijet mass distributions are driven (modulo detector
effects) by the logarithmic structure of the available
longitudinal phase space and by the value of the Pomeron
intercept, and so lead to a similar mass distribution for
Figs.~2(a) and 2(b).

The residual discrepancy between 320~fb and 50~fb may be traced,
first, to the fact that the same gap survival factor, $S^2$, is
assumed in \cite{BDPR} for LHC and Tevatron energies, whereas it
is expected \cite{KMRsoft,KKMR,GLM} that\footnote{The decrease of
$S^2$ with collider energy reflects the rise of the total
interaction cross section, and is in agreement with the D0 and CDF
data for the production of jets separated by rapidity gaps,
measured at 630 and 1800~GeV \cite{TEV630}.}
\begin{equation}
S_{\rm LHC}^2\,/\,S_{\rm CDF}^2\ \simeq\
0.4.\label{eq:Ssquaredratio}
\end{equation}
Secondly, a smaller slope, $\lambda_H = 2\lambda_{\rm dijet}/3$,
is used for Higgs, as compared to dijet, production; see eq.~(1)
of \cite{BDPR}. Neglecting the Pomeron slope, $\alpha_P^\prime$,
the cross section is proportional to $1/\lambda^2$. Finally, at
the Tevatron energy, an extra contribution to Fig.~2(b) comes from
Reggeon--Reggeon and Pomeron--Reggeon exchange interactions.
Allowing for all these effects would decrease the predicted Higgs
cross section of 320~fb by about a factor in the region of 5--10,
bringing the cross section of Ref.~\cite{BDPR} into general
agreement with our central inelastic prediction at the LHC. For
the Tevatron energy, instead of the number given in
Ref.~\cite{PP2} we have entered the later prediction of
Ref.~\cite{BDPR}.

For the central inelastic configuration, it was claimed in
\cite{BDPR} that, by tagging the outgoing protons, and by
measuring the jets accompanying the Higgs, it is possible to
obtain a good missing mass resolution for the Higgs. Unfortunately
this is only true for a centrally produced system of $M_{\rm
miss}$~close to~$M_H$, which corresponds to a very small fraction
of the events, comparable to the number for exclusive production.
Moreover, for the reasons listed above, the cross section was
overestimated.

The Pomeron--Pomeron approach of Cox et~al.~\cite{PP1} is close to
that of Boonekamp et~al.~\cite{PP2}. The main difference is that,
instead of using a soft Pomeron intercept $\alpha_P(0)=1.08$, a
larger intercept $\alpha_P(0)\sim1.2$ was used, as given by the H1
diffractive deep inelastic data. Again the prediction is
normalised to the CDF dijet data~\cite{CDFjj}. Therefore the
prediction at the Tevatron energy is reasonable. Cox
et~al.~\cite{PP1} use the same parameters for the Higgs and dijet
production amplitudes. Moreover, they use the H1 analysis of
diffractive data to specify the flux and the gluon structure of
the Pomeron. They find that their normalisation is equivalent to a
soft survival factor of $S^2\simeq0.15$ at the Tevatron. The
theoretical expectation for $S^2$ is about 0.05. This implies that
a significant part of the cross sections must come from the larger
Fig.~2(a) contributions, rather than Fig.~2(b), to compensate for
the smaller value of $S^2$.

Unlike all the previous approaches, the predictions of the Soft
Colour Interaction (SCI) model of Enberg et~al.~\cite{SCIH,SCITev}
are obtained from Monte Carlo simulations, rather than from an
analytic approach. The model assumptions on soft interaction are
implemented in PYTHIA~\cite{PYTHIA} and embody the possibility of
soft spectator rescattering and initial state QCD radiation. The
SCI model effectively incorporates the $S^2$ and $T^2$ survival
factors generated within the framework of the PYTHIA Monte
Carlo~\cite{PYTHIA}. Rapidity gaps are produced in the
model~\cite{SCIH,SCITev} by additional soft colour interactions in
the final state, which are contrived to screen the colour flow
across the gaps. The strength of these extra soft colour
interactions was tuned to reproduce the diffractive deep inelastic
data obtained at HERA. It was demonstrated that the model, with
the same parameters, describes reasonably well the single
diffractive processes observed at the Tevatron.

However, the generator was created to simulate {\em inelastic}
processes. It operates by starting from the hard subprocess and
generates the parton showers by backward evolution. The generator
never accounts for the important coherence between different
parton showers, nor for the colourless nature of the initial
particles. The incoming protons are just considered as a system of
coloured partons and only the overall colour charge is conserved.
As a consequence, the probability not to emit additional secondary
jets (and so to reproduce an exclusive process) turns out to be
negligibly small. In particular, such a generator is unable to
reproduce the elastic cross section. Originally these generators
create many secondary minijets at the parton shower stage and the
probability to screen all these minijets by colour interchange is
extremely low. Such generators were not constructed to reproduce
exclusive processes, where the colour coherence effects and
colourless nature of the incoming hadrons are important. For this
reason we believe that the extremely low limit for the exclusive
$pp\ra p + H + p$ cross section, which would follow from such an
approach, would not be trustworthy.

It is informative to note that, in our perturbative QCD
approach~\cite{KMR}, the effective Pomeron or two-gluon exchange
has relatively compact transverse size. The Sudakov-like $T$
factor occurs inside the loop integral over $Q_t$ and, in this
way, the large-size (small $Q_t$) component of the Pomeron is
strongly suppressed by QCD radiative effects. When the two-gluon
system forms a large-size colour-dipole it emits numerous
secondary gluon jets which completely fill the rapidity gap. There
is a vanishing small probability $T^2$ for the gap to survive such
emissions. The main contribution to the loop integral comes from
relatively large $Q_t$ in the region of the saddle point $Q_S$.
The value of $Q_S$ grows with both $M_H$ and the collider energy
$\sqrt{s}$. For a Higgs of mass $M_H=120$~GeV produced at the LHC,
the transverse size of the exchange is $r_P\sim1/Q_S\sim0.1$~fm.
On the contrary, in the approaches of
references~\cite{PP1,PP2,BDPR,SCITev,SCIH}, a soft large-size
Pomeron is exchanged across the rapidity gaps with transverse size
$r_P\sim r_{\rm proton}\sim1$~fm. This could cause a much stronger
Sudakov suppression if it were to be calculated by perturbative
QCD.

Another consequence of the small size of the perturbative Pomeron
concerns the validity of the $J_z=0$ selection rule for the
semi-forward hard diffractive production amplitudes. Recall that
this rule plays a crucial role in the suppression of the QCD
background\footnote{Due to the QCD factorization of soft gluon
emission (see, for example, Ref.~\cite{FACT}) the $J_z=0$
selection rule is still valid, and suppresses the $\bb$
background, even beyond leading order, arising from events where
the $\bb$ pair is accompanied by one or more soft
gluons~\cite{DKMOR}. Hence the QCD-induced $\bb$ background is
expected to be suppressed both for the exclusive process and for
low mass central-inelastic production where the missing mass
$M_{\rm miss} \equiv M_{PP}$ is close to
$M_H$.}~\cite{Liverpool,DKMOR}. In the exact forward direction,
$J_z=0$ by virtue of angular momentum conservation. However,
violation of this rule can come from orbital angular momenta,
$r\,p_{it}$, where $p_{it}$ is the transverse momentum of the
leading proton and $r\sim 1/Q_t$ is the transverse size of the
Pomeron. Therefore the admixture of the $|J_z|=2$ state is
strongly suppressed for the small-size Pomeron-exchange occurring
in the exclusive amplitude~\cite{KMRmm}. On the other hand, for
C-inelastic production, where the $T$ factor becomes inactive and
we deal with a large-size Pomeron, we lose the $J_z=0$ selection
rule and, as a result, have a much larger background. The same is
true for Monte-Carlo-based models. The soft colour interaction,
which screens the colour across the gap, takes place at large
distances and therefore we have no $J_z=0$ selection rule. So the
expected signal-to-background ratio is small.

\section{R\'esum\'e}

We compiled a representative range of different predictions of the
cross sections for diffractive production of a Higgs boson of mass
about 120~GeV at the Tevatron and LHC. We critically compared the
wide range of predictions and explained the origin of the
differences. In summary, the wide spread of predictions occurs
either because different processes have been considered or because
important effects have been neglected.

The cross sections for inclusive and central inelastic diffractive
Higgs production are larger than for exclusive production.
However, for these non-exclusive processes it is hard to suppress
the QCD $\bb$ background and the signal-to-background ratio is
small. Second, we cannot improve significantly the accuracy of the
measurement of the mass of the Higgs boson by tagging the forward
protons and measuring the missing mass.

On the other hand, the cross section for exclusive diffractive
production is known with sufficient accuracy to be sure that this
channel can be used to play an important role in Higgs
detection\footnote{Unfortunately, the cross section is too low for
this method to be used at the Tevatron.} via $H\ra\bb$ at the LHC,
provided that forward proton taggers are installed. The mass of
the Higgs could then be accurately measured by the missing-mass
method, $\Delta M_{\rm missing}\simeq1$~GeV~\cite{DKMOR}.
Moreover, the leading order $\bb$ background is strongly
suppressed by a $J_z=0$ selection rule.

Details of the calculation of the $pp\ra p+H+p$ exclusive Higgs
production cross section are given in Ref.~\cite{INC}. The cross
section is predicted to be 3~fb at the LHC, with a factor of two
uncertainty~\cite{DKMOR}. The main sources of the $\bb$ background
are, at leading order, caused by gluon jets being misidentified as
a $\bb$ pair, by a $J_z=2$ admixture due to non-forward protons
and by a $J_z=0$ contribution arising from $m_b\neq0$. Also there
is a background contribution from $\bb g$ events in which the
emitted gluon is approximately collinear with a $b$ jet. These
backgrounds were considered in detail in Ref.~\cite{DKMOR},
leading to a signal-to-background ratio of about 3. Note that in
\cite{DKMOR} only the $gg\ra\bb g$ hard subprocess was considered
at NLO, and radiation for the spectator, screening gluon was not
discussed. However, this latter process is numerically small
because of the additional suppression of colour-octet $\bb$
production around $90^\circ$; rotational invariance around the $b$
quark direction causes the cross section to be proportional to
$\cos^4\theta$ in the $\bb$ c.m. frame~\cite{BORDEN}.

We may summarize the exclusive diffractive Higgs signal ($pp\ra p
+ H + p$ with $H\ra\bb$) by the following example. Consider the
detection of a Higgs of mass 120~GeV with an integrated luminosity
of 30~fb$^{-1}$ at the LHC. When account is taken of the
efficiencies associated with proton tagging and with the
identification of $b$ and $\bar{b}$ jets, and allowance is made
for the polar angle cuts and the $H\ra\bb$ branching ratio, then
the original $(\sigma=3~{\rm fb})\times({\cal L} = 30~{\rm
fb}^{-1})=90$ events is reduced to an observable signal of 11
$H\ra\bb$ events, with a background of 4~\cite{DKMOR}.

We stress that the predicted value of the exclusive cross section
can be checked experimentally. All the ingredients, except for the
NLO correction to the $gg\ra H$ vertex, are the same for our
signal as for exclusive double-diffractive dijet production,
$pp\ra p+ {\rm dijet} + p$, where the dijet system is chosen in
the same kinematic domain as the Higgs boson, that is
$M(jj)\sim120\:{\rm GeV}$ \cite{INC,KMR}. Therefore by observing
the larger dijet production rate, we can confirm, or correct, the
estimate of the exclusive Higgs signal.

\section*{Acknowledgements}

We thank Brian Cox, Albert De Roeck, Rikard Enberg, Jeff Forshaw,
Aliosha Kaidalov, Genya Levin, Leif L\"onnblad, Risto Orava, Robi
Peschanski, Christophe Royon, Torbj\"orn Sj\"ostrand and Michael
Spira for useful discussions. One of us (VAK) thanks the
Leverhulme Trust for a Fellowship. This work was partially
supported by the UK Particle Physics and Astronomy Research
Council, by the Russian Fund for Fundamental Research (grants
01-02-17095 and 00-15-96610) and by the EU Framework TMR
programme, contract FMRX-CT98-0194 (DG 12-MIHT).

\end{document}